\documentclass[a4paper,aps,pre,10pt,floatfix,twocolumn,showpacs,superscriptaddress,nofootinbib]{revtex4-1}

\usepackage{subfigure}
\usepackage{amssymb}
\usepackage{amsfonts}
\usepackage{amsmath}
\usepackage{amsthm}
\usepackage{epsfig}
\usepackage{graphicx}
\usepackage[normalem]{ulem}
\usepackage[usenames,dvipsnames]{color}
\usepackage[latin1]{inputenc}
\usepackage{comment}

\begin{document}

\title{Fractal Dimension and Universality in Avascular Tumor Growth}

\author{Fabiano  L. Ribeiro \footnote{fribeiro@dex.ufla.br}}

\author{Renato Vieira dos Santos \footnote{\url{renato.santos@dfi.ufla.br}}}

\author{Ang\'{e}lica S. Mata \footnote{\url{angelica.mata@dfi.ufla.br}}}

\affiliation{Departamento de F\'isica, Universidade Federal de Lavras, 37200-000, Lavras, MG, Brazil}

    \begin{abstract}

The comprehension of tumor growth is a intriguing subject for scientists. New researches has been constantly required to better understand the complexity of this phenomenon. In this paper, we pursue a physical description that account for some experimental facts involving avascular tumor growth. We have proposed an explanation of some phenomenological (macroscopic) aspects of tumor, as the spatial form and the way it growths, from a individual-level (microscopic) formulation.
The model proposed here is based on a simple principle: competitive interaction between the cells dependent on their mutual distances.
As a result, we reproduce many empirical evidences observed in real tumors, as exponential growth in their early stages followed by a power law growth. The model also reproduces the fractal space distribution of tumor cells  and the universal behavior presented in animals and tumor growth, conform  reported by West, Guiot {\it et. al.}\cite{West2001,Guiot2003}.
The results suggest that the  universal similarity between tumor and animal growth 
comes from the fact that both are described by the same growth equation - the Bertalanffy-Richards model -  even they does not necessarily share the same biophysical properties.

\end{abstract}


\maketitle


\section{Introduction}\label{introduction}
\label{intro}


The understanding of tumor growth has for long been regarded as a subject of outstanding interest in many fields such as biology, oncology, mathematics and physics \cite{history1,math_oncology,dynamics_cancer}. 
Empirical observations lead us to believe that the growth of tumors are ruled by general growth laws which can be represented by differential equations \cite{model_muddle}. The advantages of these representations are uncountable. These models can be useful to estimate the progress of the tumor and, consequently, to  
determine the frequency for performing exams such as mammography screening or
to control the success of some therapy \cite{oncolydrug,naturegompertizian,thyroidalnodules,tumorxenografts}.
Furthermore, experimentalists are becoming increasingly convinced that 
mathematical modeling can clarify and interpret many experimental findings.

Even after decades of study, researchers are still attempting to answer the main question with regard to this topic: how the tumor
increases over time as the disease develops? 
Even though this point may seem simple, the basic mechanisms underlying growth of tumors are still not clear enough.
The Gompertz curve, one of the most important models used for such studies, has been able to describe many animal and human data \cite{West2001, model_muddle, wheldon, laird}.
On the other hand, the Bertalanffy-Richards model \cite{Yuancai1997, Review1949,richards1959}, 
which has Gompertz, Verhulst and exponential growth models as particular cases \cite{model_muddle, tests_tumors_models,Martinez2008,Ribeiro2016},  provides accurate description of tumors growth as well \cite{comparison_tumors_models}. 

Within this context, we are interested in characterizing avascular tumor \cite{thecell} to better understand it develpoment. Although avascular growth correspond only to the initial stage of tumors, the knowledge of 
this phase is quite important since most empirical evidences are based on avascular tumor spheroids {\it in vitro}  \cite{spheroids_vitro,vitro_cancer_3D,cancer_invitro}. This kind of experiments are easily manipulated as opposed to {\it in vivo} ones.

Our current research, supported by previous works \cite{Ribeiro2015b,mombach2002}, proposes a  microscopic model  to describe tumors growth using just a few physical principles. However, it is able to clarify empirical evidences and reproduces some widely known models under certain conditions. Our model was formulated based on distance-dependent competitive interactions between cells. Even considering just this very simple interrelation, our approach is able to interpret (at microscopic level) the phenomenological  Bertalanffy-Richards model and consequently, the Gompertz curve. We also explain the fractal distribution of the tumor cells, verified by empirical studies
\cite{Article1997,Baish2000a}.

\begin{figure}[t]
 \centering
\includegraphics[width=\columnwidth]{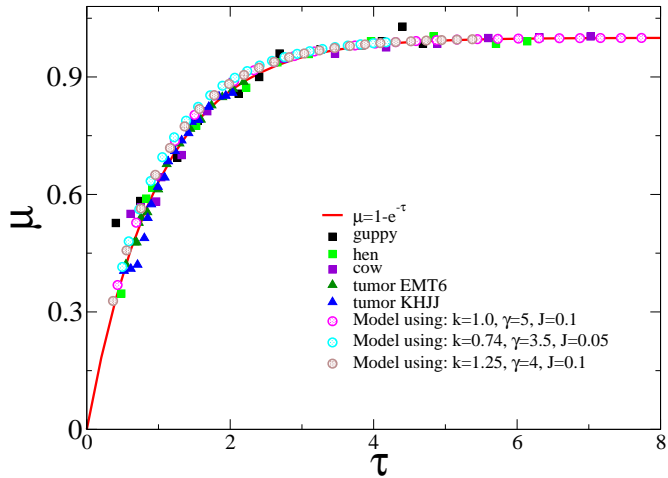} 
\caption{\label{fig_universal} The universal growth law of the dimensionless mass ($\mu$) as a function of the dimensionless time ($\tau$). Completely different animals, tumors and data from the model proposed (see section~(\ref{sc5})), regardless the parameter values, collapse in the same universal curve.}
\end{figure}

In addiction, the model presents a widely robust behavior. This mean that no matter the parameters values of the model, all the setting always collapse in the same curve. And this curve is exactly the same universal curve that describes tumor and animal growth, as shown in figure~(\ref{fig_universal}). 
 West {\it et. al.} \cite{West2001} explains that the universal behavior observed in animal growth is due to the similar way that the organisms allocate energy to growth and to maintenance. Although this pattern is also observed for tumors, the biological mechanisms behind it is not clear enough. 
However, our model is able to explain this phenomenon, because we show that the universality is a key feature, since it just comes from two microscopic principles: competition and self-replication.
 
 It is well known that universality is observed in many physical and biological systems \cite{Kadanoff2000,Yeomans1992,Stanley1987}. 
 So, our model will be very useful to investigate universality and common patterns in population growth in general (tumor, animal and any other one). This subject has drawn attention of many theoretical \cite{Martinez2008,Chester2011,Cabella2011, Cabella2012a,Ribeiro2014,Ribeiro2015c, Kuehn2010, Strzalka2008, Strzalka2009, Torquato2011} and empirical \cite{West2001,Guiot2003, Guiot2006, savage} researches in the last decades.

This paper is organized as follows: section~(\ref{empirical}) gives a brief description of some empirical evidences that are considered common features in the most of 
avascular tumors. 
In the section~(\ref{sectio_micro}) our mathematical model to tumor growth is presented. 
In section~(\ref{sc3}) we show the concept of optimal fractal dimension, whose consequences for tumor 
growth will also be explained.
In section~(\ref{sc4}) we get the Bertalanffy-Richards model using first principles and we also obtain the Gompertz model as a particular case of our approach. 
In section~(\ref{sc5}) we show that our model is able to explain why tumors, as well as animal growth, also follow the universal behavior. 
Both are described by the same mathematical - Bertalanffy-Richards -  model,    even though they do not necessarily follow the same biological principles.  Finally, in the last section we present our conclusions. The appendix~(\ref{apendice_modelo_west}) is just a very short review of the West model \cite{West2001}.

\section{Empirical Evidences}
\label{empirical}

\begin{figure*}[t]
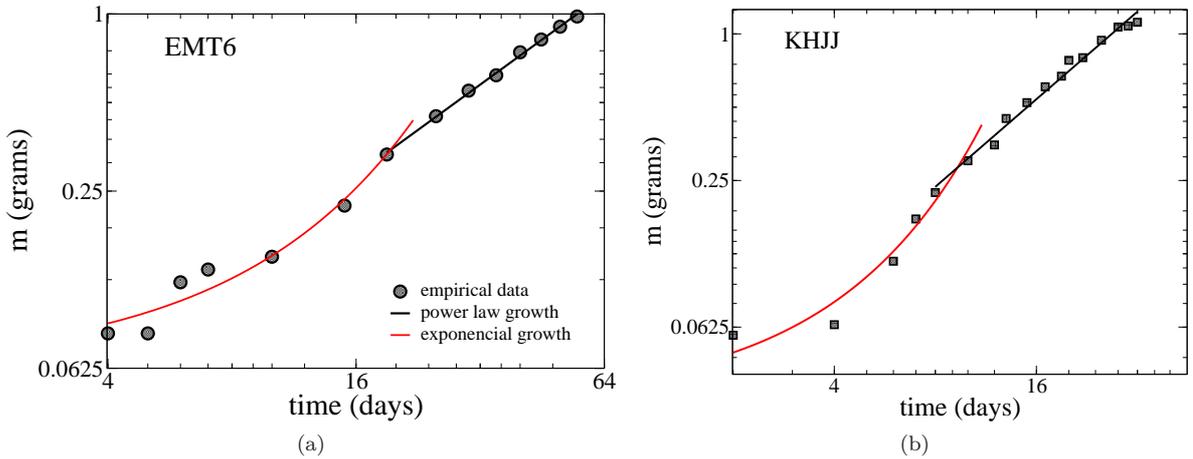

\centering
\subfigure[]{\includegraphics[scale=0.295]{EMT6_edited2.eps}}\quad
\subfigure[]{\includegraphics[scale=0.265]{KHJJ_edited2.eps}}
\caption{\label{Fig_dados} 
Empirical data of growth  (the dots) of two types of tumors (a) EMT6 and (b) KHJJ implanted in mice and rats. These data was extracted directly from the reference \cite{Herman2011} and  represent the usual growth behavior in tumors: exponential growth in the first stage followed by a power-law growth.}
\end{figure*}

Many empirical works \cite{comparison_tumors_models,Modelling2003,math_tumor_model, prospective} suggest that the total mass $m$ forming the tumor increases with time $t$ obeying two different regimes: an initial exponential growth followed by a power-law one.
That is, after the early exponential stage, we have
\begin{equation}
m(t)\sim t^{\alpha},
\end{equation}
where $\alpha$ is a exponent which depends on the tumor type and the micro-environment conditions  \cite{Modelling2003,fractals_cancer, history2}.
The data presented in Fig.~(\ref{Fig_dados}), reproduced from \cite{Modelling2003}, show this typical time evolution of tumor size. 
 Moreover, some experiments \cite{Modelling2003, Freyer1985}  also show that the radius of the solid tumor, $R_{max}$, grows linearly with time ($R_{max} \sim t$). Consequently, the spatial size of the tumor scales with its mass as
\begin{equation}
 R_{max} \sim m^{1/\alpha}.
\end{equation}

Previously, tumor growth was characterized as chaotic \cite{history2,Guiot2006}. But nowadays it is known that fractal geometry is more suitable to quantify the morphological characteristics of solid tumors \cite{fractals_cancer,savage,DOnofrio2009a}. 
This property has also been used in the diagnosis concerning the tumor malignancy. Some studies suggest that higher fractal dimension implies greater tumor aggressiveness \cite{dobrescu2002using, tumourshape, fractalpathology, fractalpatterns}.

Another empirical aspect about tumors is that its growth seems to present the same universal pattern observed in animal growth  \cite{Guiot2006,Guiot2003}.
West {\it et. al.}~\cite{West2001} first proposed that the growth of different species (mammals, birds and fish) collapse in the same universal curve. After it was also verified for tumor growth (in \textit{vivo}, \textit{in vitro}, and  in clinical practice)  \cite{Guiot2003,Guiot2006,savage}.
Figure~(\ref{fig_universal}) summarize such finds, showing the 
plot of the dimensionless mass ($\mu$) as a function of the dimensionless time ($\tau$), for completely different kind of animals (cow, chicken, and guppy), and for two different tumors (the ones described in the figure ~(\ref{Fig_dados})). 
All of them collapsing in the same universal curve: $\mu = 1- e^{-\tau}$ (more details in appendix~(\ref{apendice_modelo_west})).

The universal behavior of animal growth is because all species 
allocate energy for growth and maintenance in a similar way  \cite{West2001,Herman2011,savage}. However, for tumors, this universal behavior is still 
not clear enough. 
In the present work, we will show that 
self-replication and competition among the cells
can also generate the  universal growth pattern.

In order to corroborate this empirical evidences 
we propose a simple microscopic model, based 
on the distance-dependent interaction between cells living in a competitive environment. 
This model explains the following properties of solid-tumor growth: 
\begin{enumerate}
\item  exponential growth in early time;
\item  power law growth for large enough time; 
\item  diameter follows a power law with the number of cells;
\item fractal-like structure; 
\item universal behavior.  
\end{enumerate}
The details of the model is presented in the next section.

\section{The microscopic model}\label{sectio_micro}

 We have formulated a simple mathematical model based on four basic ``principles'': 

\begin{enumerate}

\item The cells compete by resources in their micro-environment;

\item The cells moves in order to minimize the competitive interaction. 

\item The replication rate of each cell is affected by its competitive interaction with the other ones;

\item The intensity of the competitive interaction decays with the distance between the cells;

\end{enumerate}

We will show in the next sections that these basic premises are enough to explain the tumor growth empirical evidences which were reported on previous section. 
In order to put these principles in a mathematical way, consider a \textit{interaction function}  $f(r)$  which represents the effects - \textit{the field} - perceived by a
single cell, say $i$, due to the presence of another cell at distance $r$. Consider also  that the population of cells is spatially distributed according to the density function  $\rho(\mathbf{r})$, where $\mathbf{r}$ is the \textit{position vector}. Then the total interaction field affecting the cell $i$ is 

\begin{equation}\label{eq_Ii}
 I_i = \int_{V_D} \rho(\mathbf{r}) f(r) d^D\mathbf{r}. 
\end{equation}
Here, $D$ (integer) is the Euclidean dimension, $d^D\mathbf{r}$ is the hyper-volume element, and $V_D$ the hyper-volume in which the population is immersed.

It is quit plausible that the interaction between two cells decay with distance, and therefore, a reasonable
choice for the interaction function is 
\begin{equation}
 f(r) =  
 \left\{ \begin{array}{ll}
\frac{1}{r^{\gamma}}& \textrm{  if $r>2 r_0$ } \\
\frac{1}{(2r_0)^{\gamma}} &  \textrm{ otherwise} 
\end{array} \right. ,
\label{decay}
\end{equation}
where $\gamma$ is the decay exponent, and $r_0$ is the diameter of the cell. 
The hypothesis~(\ref{decay}) has been used recently to describe population cell growth \cite{mombach2002,Ribeiro2015b}.  
Consider hereafter, by convenience, $r_0 = 1/2.$

Based on the empirical evidence described in subsection (\ref{empirical}), let's suppose the population evolves forming a  structure with dimension $D_f \le D$. 
While the number of cells scales as $r^{D_f}$, the hyper-volume in which the cells are inserted scales as $r^D$ \cite{falconer}.  
This fact allows us to write the density of cells as 
\begin{equation}\label{Eq_rho}
 \rho(\mathbf{r}) = \frac{\textrm{Number of cells}}{\textrm{Volume}} = \rho_0 \frac{r^{D_f}}{r^D} = \rho_0 r^{D_f - D}, 
\end{equation}
where $\rho_0$ is a constant.

With these assumptions one can solve~(\ref{eq_Ii}) considering  \textit{periodic boundary conditions}, and using 
$d^D\mathbf{r} = r^{D-1}dr\, d\Omega_D$,  
where $d\Omega_D$ 
is the solid angle. These considerations allows us to write the interaction field as (see details in ref. \cite{Ribeiro2015b})

\begin{equation}\label{eq_Ii2}
 I_i = I (D_f, N) =  \frac{\omega_D}{D_f}
 \ln_{(\beta -1)} \left( \frac{D_f}{\omega_D} N \right)  + \frac{\omega_D}{D_f},
\end{equation}
where 
\begin{equation}\label{Eq_beta1}
\beta \equiv 2- \frac{\gamma}{D_f},
\end{equation}
and   $\ln_{\beta-1} (x) \equiv (x^{\beta-1}-1)/(\beta -1)$ is the \textit{generalized logarithm} (see \cite{tsallis2009}). The particular case $\beta \to 1$ (or $\gamma \to D_f$) leads to the \textit{natural logarithm}. 
Some recent works \cite{Cabella2011,Ribeiro2015b,Ribeiro2015c} suggested that generalized growth models can be written in terms of \textit{generalized logarithm}, which allows a easier 
algebraic treatment.
In Eq.~(\ref{eq_Ii2}) we also introduced $\omega_D \equiv \rho_0 \int d\Omega_D$ which is a constant 
that depends only on the dimension $D$.
In the case $N \to \infty$, the interaction field~(\ref{eq_Ii2}) diverges if $\gamma < D_f$, which implies the existence of long range interactions between cells, and converges if $\gamma > D_f$, a short range interaction between cells. 
We will restrict this work to the short range interaction situation, that is,  $\gamma > D_f$, which means that the cells interact only with their closer neighbors.
In this case, the periodic boundary conditions simplify the problem and it is good enough to describe realistic situations.

Note that the field given by Eq.~(\ref{eq_Ii2}) does not depend on the index $i$, that is, it is the same 
for all the individuals of the population. It is a consequence of the  periodic boundary conditions \cite{Ribeiro2015b}  and the self-similarity of the fractal structure formed by the tumor \cite{falconer}.
The result~(\ref{eq_Ii2}) means that all cells feel the same influence from their neighbors cells.  In fact this influence depends only on the fractal dimension and the size of the population, that is  $I_i = I(N, D_f)$. 

In section~(\ref{sc3}) it is showed that the population reaches a fractal space distribution  when the cells move in order to minimize the competitive interaction. That is, the model is able to explain,  by a microscopic approach, the spacial distribution observed in real tumor.

As a first application of this model, one can use it to 
predict how the  diameter of the tumor,  say $2 R_{max}$, behaves with the increase of the number of cells $N$. 
Considering that  $N = \int_{\textrm{all space}} \rho(\mathbf{r}) d^D\mathbf{r}$ and the density given by Eq.~(\ref{Eq_rho}),  one has 
$N = \rho_0 \int d\Omega_D \int_{0}^{R_{max}} r^{D_f-1} dr$ 
and, consequently,
\begin{equation}\label{Eq_Rmax}
 R_{max} \sim (D_f N)^{\frac{1}{D_f}}. 
\end{equation}
This result is in accordance with the empirical fact that the diameter of the tumor follows a power law with the number of cells \cite{Modelling2003,Freyer1985}.

\section{Optimal Fractal Dimension}
\label{sc3}

\begin{figure}[htb!]
\centering
\includegraphics[scale=0.2]{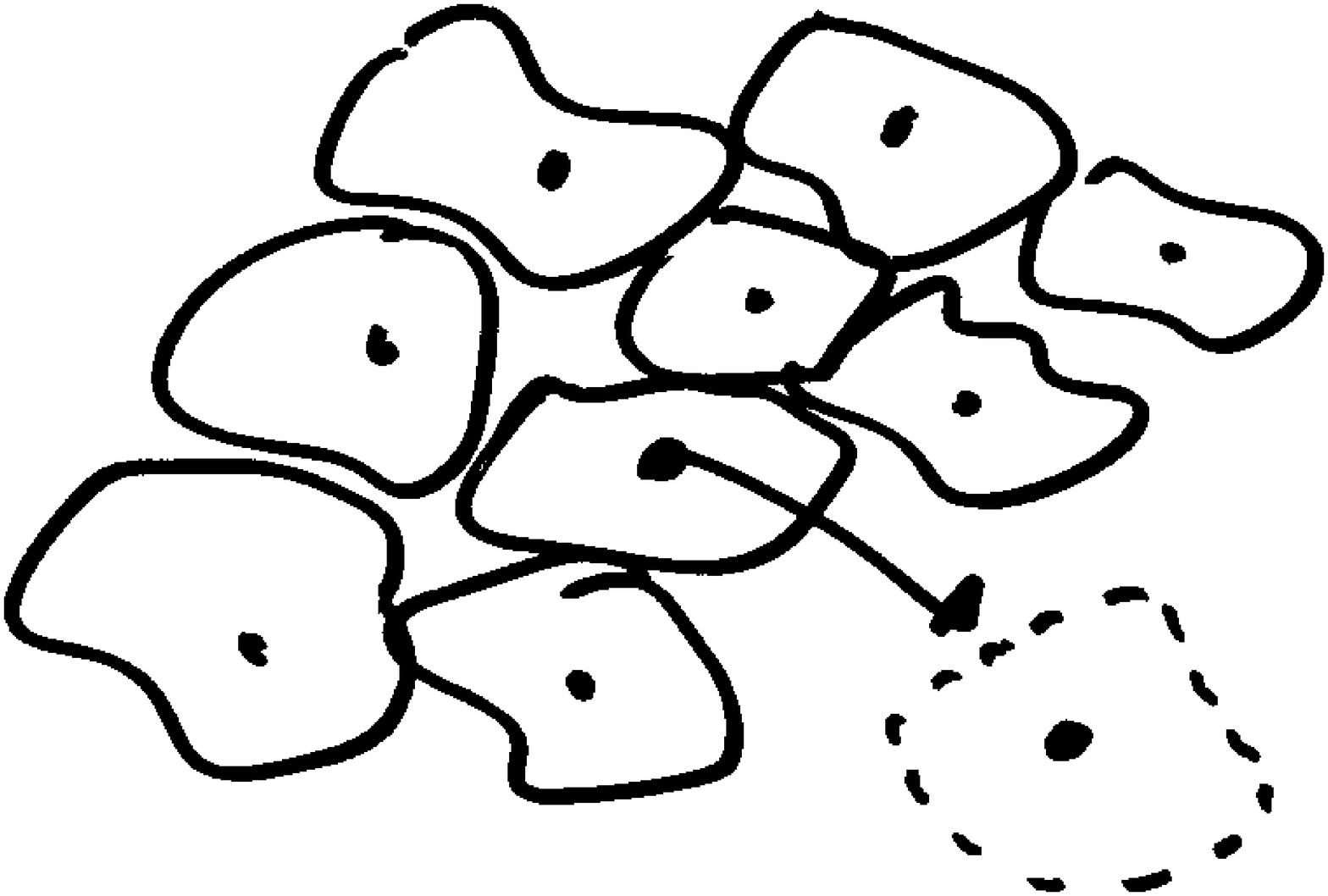}
\caption{ \label{Fig_cells} Schematic representation of the cellular dynamics. In competition, each cell tries to move in a direction that minimize the influence of the other cells. In the case represented by this figure, the cell moves in the direction that 
minimize the ``pressure'' from the other cells.}
\end{figure}

During growth, the cells compete by available resources in their micro-environment. It means that each cell must to move in order to minimize the competitive influence  from the other cells. Fig.~(\ref{Fig_cells}) shows schematically  the movement of a single cell to minimize the ``pressure'' from its neighbors. 
If  all cells perform such a movement, the spatial distribution of the population will change 
adaptively. Consequently the (fractal) dimension of the structure formed by the population will change.

\begin{figure}[htb!]
\centering
\includegraphics[scale=0.25,angle=-90]{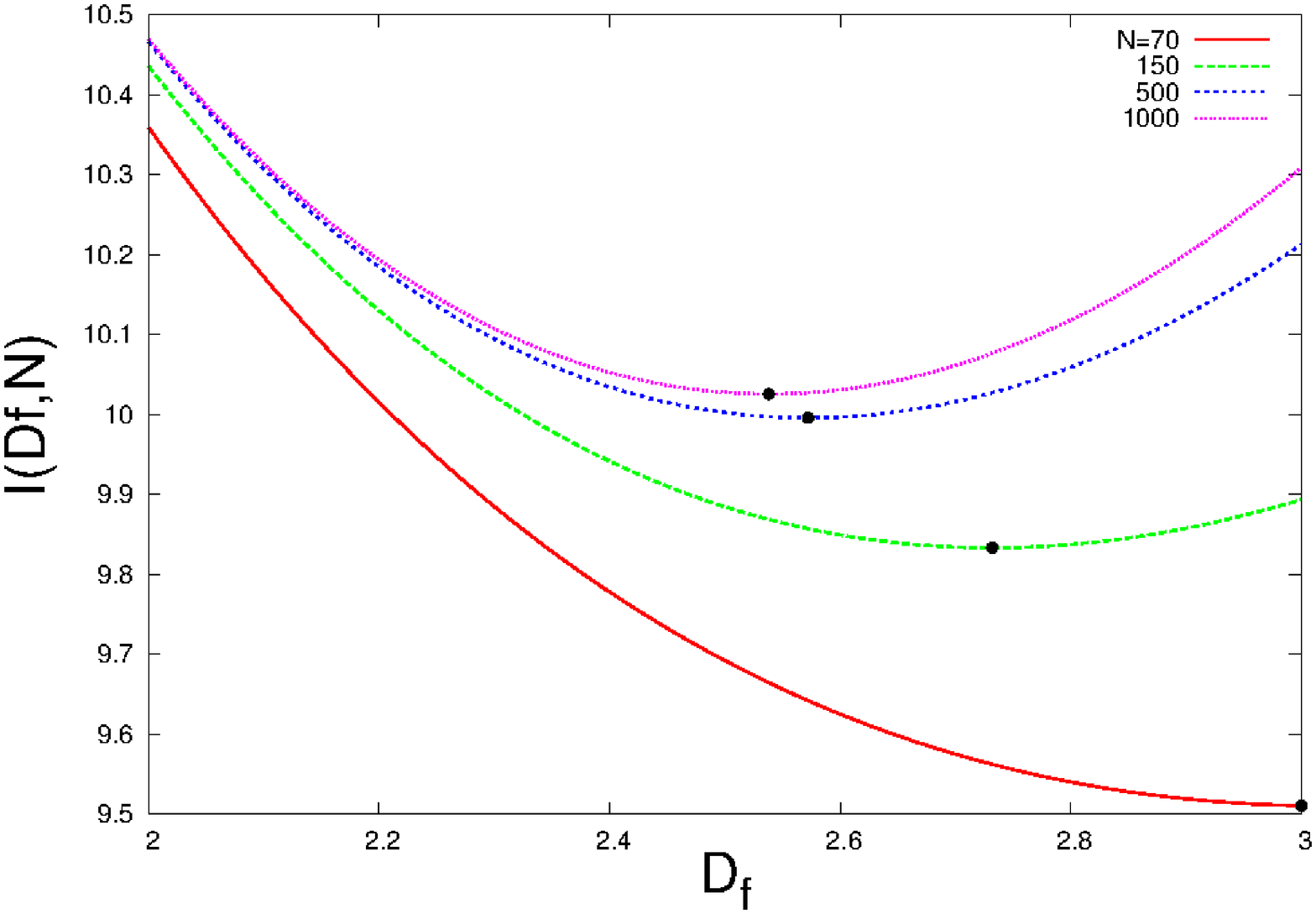}
\vskip 0.5 cm 
\includegraphics[scale=0.24]{Df_estrela-x-N}
\caption{\label{Fig_IixDf}  
(a) Plot of the field $I(N, D_f)$, given by Eq.~(\ref{eq_Ii2}), as a function of the fractal dimension. The plot were performed for some fixed values of population size and using $\gamma=5$.  The black dots represent the minimum value of $I(N,D_f)$
in relation to the fractal dimension, which leads to $D_f^{opt}$.  
When the population is small, the plot suggests that the population tends to be compact, that is $D_f^{opt} = 3.$ However, when the population becomes sufficiently large,
the optimal dimension is fractal. (b) Plot of the fractal dimension which minimize the interaction field ($D_f^{opt}$) as a function of the population size. The optimal fractal dimension decreases monotonically as the population increases, and more slowly for sufficient large population.
}
\end{figure}

The interaction field generated by the cells presents two extreme values: when  $D_f \to 0$ and $D_f \to 3$. It is because if $D_f \to 0$ the population tends to concentrate in a single point (see Eq.~(\ref{Eq_Rmax}): $R_{max} \to 0$  when $D_f \to 0$),  generating a high interaction field. On the other extreme, if $D_f \approx 3,$ and $N$ is sufficiently large, the population is 
compacted (high density), which also result in a strong interaction field. Then, there is an \textit{optimal fractal dimension} around these two extremes, that we will call $D_f^{opt}$, which minimize the competitive influence among the cells. 
That is,  $D_f^{opt}$ is the fractal dimension which minimize the field~(\ref{eq_Ii2}) experienced by each cell, that is obtained by the condition

\begin{equation}
 \frac{\partial }{\partial D_f} I(N,D_f) \Bigg|_{D_f=D_f^{opt}} = 0.
\end{equation}

Such ideas are illustrated quantitatively in Fig.~(\ref{Fig_IixDf}(a)), which presents the plot of the interaction field $I(N, D_f)$, given by Eq.~(\ref{eq_Ii2}),  as function of the fractal dimension of the population, keeping the population size fixed. By this plot it is possible to see that the interaction field always has a minimum (in $D_f = D_f^{opt}$) if the population is large enough. 
If $N$ is small $D_f^{opt}=3$ (maximally compacted population),  but for a sufficiently large population, the optimal dimension is fractal.

Our microscopic model clarifies the role of the fractal geometry of tumor growth since this structure emerges spontaneously as a response of the cells to minimize competitive pressure. 
In the next section we present the population growth process in this competitive context. We will see that the model proposed presents an early exponential stage, followed by a power-law growth regime, in accordance with the empirical evidences over again.


\section{Population Growth}
\label{sc4}


Our model presents two time scales.  The first one is the time that the cells take to reach the optimal fractal dimension arrangement. The second time scale, represented by $t$,  is  measured in generations. 
At the end of each generation the cells reproduce. Consider that the time between two consecutive generations is sufficiently large to allow population to reach the optimal spatial distribution, and consequently achieves $D_f^{opt}$,  before reproduction. 
Then, at the moment of the reproduction, the field felt by each cell is $I^{opt}(N) \equiv I(N, D_f=D_f^{opt})$.

In a \textit{competitive} context, the interaction field exercise a inhibitory behavior in the reproduction of the cells \cite{mombach2002,DOnofrio2009a,Ribeiro2015b}. 
In this way, it is quite reasonable to consider that the \textit{replication rate} $R_i$ of the $i$-th cell is given by
\begin{equation}
R_i = k - J I^{opt}.  
\end{equation}
{This equation says that the reproductive capability of the cells depends in part on an inherent property - given by the 
\textit{intrinsic replication rate} $k$ -,  and in part on the influence (inhibitory) of the other cells of the population - given by $J I^{opt}$.
The parameter  $k$ is identical, by definition, to all cells, while $J I^{opt}$ represents the \textit{rate of competition}.  
The parameter $J>0$ measure the intensity of the inhibition (competition).
The case $J<0$ provides us a different approach than we are interested, which means cooperation between cells, and was already discussed in  previous works \cite{Ribeiro2015b,Ribeiro2015c,DosSantos2015a}.

Since $\Delta t R_i$ is the number of daughters that the cell $i$ generates in a time interval $\Delta t$, 
the update of the size of the population
in this period is 

\begin{equation}\label{Eq_recorrencia}
 N(t+\Delta t) = N(t) + \Delta t  \sum_{i=1}^N R_i.
\end{equation}
In the limit $\Delta t \to 0$, one has an Ordinary Differential Equation (ODE): 
\begin{equation}
 \frac{dN}{dt} = N \Big( k - J I^{opt} \Big).
\end{equation}
Introducing the result given by Eq.~(\ref{eq_Ii2}), one gets
\begin{equation}\label{EDO_N}
 \frac{dN}{dt} = c N^{\beta^{opt}} - b N 
\end{equation}
where 
\begin{equation}\label{Eq_b}
 b \equiv  \frac{J \omega_D \gamma}{D_f^{opt}(\gamma - D_f^{opt})} - k, 
 \end{equation}
 \begin{equation}
  c \equiv \frac{-J}{\left( 1- \frac{\gamma}{D_f^{opt}} \right)} \left( \frac{D_f^{opt}}{\omega_D} \right)^{-\frac{\gamma}{D_f^{opt}}},
 \end{equation}
and
\begin{equation}\label{Eq_beta}
\beta^{opt} \equiv 2- \frac{\gamma}{D_f^{opt}}.
\end{equation}

Assuming that each cell of the population has the same mass $m_c$, the total mass of the tumor at time $t$ can be written as $m(t) = m_c N(t)$. So, Eq.~(\ref{EDO_N}) becomes 
\begin{equation}\label{EDO_bertalanfy}
 \frac{dm}{dt} = a m^{\beta^{opt}} - bm,
\end{equation}
with
\begin{equation}\label{Eq_a}
 a \equiv  \frac{ - J m_c^{\frac{\gamma}{D_f^{opt}} -1} }{1- \frac{\gamma}{D_f^{opt}}}
\left( \frac{D_f^{opt}}{\omega_D} \right)^{-\frac{\gamma}{D_f^{opt}}}.
\end{equation}
Note that this model reproduce the Bertalanffy-Richards model when the parameter $D_f$ is constant.
This particular case is discussed in the section~(\ref{section_bertalanffy}).

As $\beta^{opt}$ is not fixed (it depends on $D_f^{opt}$, which in turn depends on the population size), analytical solution of Eq.~(\ref{EDO_bertalanfy}) is difficult to be  obtained. 
However, the dynamics of the model can be  investigated solving numerically the  recurrence relation given by Eq.~(\ref{Eq_recorrencia}). 
This computational solution is presented in Fig.~(\ref{fig_mxt}), which shows that  the population size (the mass) of the tumor 
growths exponentially at the beginning, and then pass through a power law regime before saturates or blow up. 
If the population growth according to one (saturation) or other (blow up) regime depends on the value of the self-replication rate. 
The value of $k$ that divides these two regime  is 
$k=k_{power}$, where

\begin{equation}\label{eq_kpower}
k_{power} \equiv  \frac{J \omega_D \gamma}{D_f^{opt} (\gamma- D_f^{opt})}
\end{equation}
is obtained taking $b=0$ in Eq.~(\ref{Eq_b}) (more details in the section~(\ref{section_bertalanffy})).

When $k=k_{power}$ the population grows purely in a power law regime, described by (see section~(\ref{section_bertalanffy}))
\begin{equation}\label{Eq_power_m}
 m(t) \sim t^{ \frac{D_f}{\gamma - D_f} }. 
\end{equation}
In fact, the power law growth happens because the replication rate $R_i$ decreases as a power law in this case (see Fig~(\ref{fig_mxt}-b)). 
The population saturates when $k<k_{power}$ because in this case the replication rate go to zero as the time evolves. And finally, the population blow up (exponentially) when $k>k_{power}$ because, in this case,  the replication rate $R_i$ converge to a constant (greater than zero)\footnote{Constant replication rate conducts to a Malthusian growth.}. 

What is important to note in these numerical results is that no matter what the value of the intrinsic replication rate is, the dynamics of the population always pass through a power law growth regime, in accordance with what is observed in empirical tumor growth.  According to the numerical results,  the time that the population growth stays on this regime depends on the intrinsic replication rate of the cells. If $k \sim k_{power}$  then the population stay a large period on a power law growth, but if $k$ is very different from $k_{power}$ the population stays only a small period on this regime   
(see figure~(\ref{fig_mxt})). 

The dynamics of the fractal dimension $D_f^{opt}$ is also presented in Fig.~(\ref{fig_mxt}). Note that this dimension  decreases rapidly when the population is small and then saturates, for $N$ sufficiently large, to a value that we will call  $D_f^{conv}$ (convergence value).  When $k\ge k_{power}$, then   $D_f^{conv} \to   \gamma/2$.

\begin{figure*}[htb!]
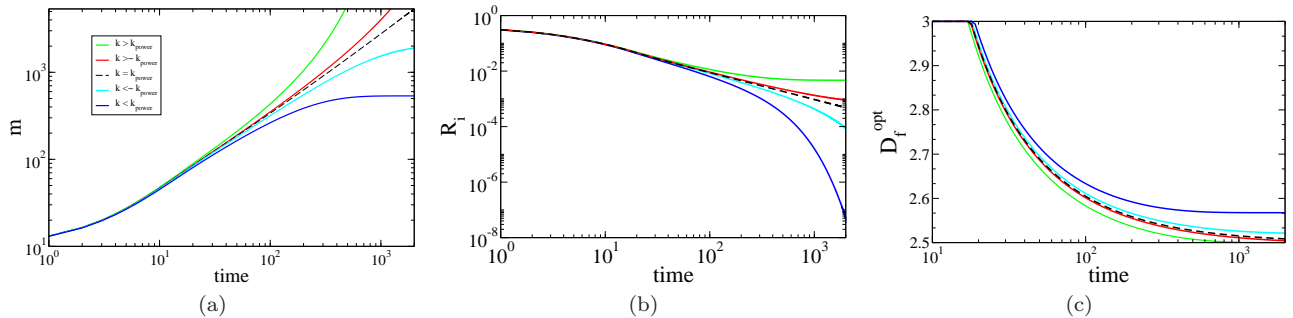

\subfigure[]{\includegraphics[scale=0.20]{mxt2_edited2.eps}}\quad
\subfigure[]{\includegraphics[scale=0.202]{Rixt_edited.eps}}\quad
\subfigure[]{\includegraphics[scale=0.202]{Dfxt_edited.eps}}
\caption{\label{fig_mxt} 
Dynamics generated by the microscopic model: (a) mass, (b) replication rate, and (c) optimal fractal dimension. 
It was used some possible values of the intrinsic replication rate $k$, and keeping fixed the parameters  $J=0.1$,  $\gamma=5$, and  $m_c=1$.
The curves were getting solving the relation of recurrence given by Eq.~(\ref{Eq_recorrencia}). When $k>k_{power}$ the population diverges; when $k = k_{power}$ the population (or its mass) growth as a power-law; and when  $k<k_{power}$ the population saturates.
If $k \sim k_{power}$,  the population pass a long period in a power-law growth regime. 
The population growth as a power law regime  because the replication rate $R_i$ also decreases as a power law. 
The optimal fractal dimension 
decreases rapidly when the population is small and then saturates (to 
$D_f^{cov}$) when the population is sufficiently large.}
\end{figure*}

In a real situation,  $k<k_{power}$ must be obeyed (otherwise the population blows up), and then the population must saturates in a sufficiently large time. 
However, as discussed in \cite{Guiot2006,savage,Guiot2003}, 
the real tumor does not saturate because before this the patient is died or the tumor cells start the vascularization and, consequently, the metastasis. 
Nevertheless, our microscopic model is good enough to describe the  power law growth phase of real tumors.  

The merit of the model proposed here is that it explain,  with few principles, the spatial distribution and the growth process of real tumor. 
It is important also to empathize that the model is built using a microscopic (individual-level) approach,  and not  a phenomenological (macroscopic) perspective, as
is usually done \cite{ history1,model_muddle,Ribeiro2015b,history2,history3}.

\subsection{Connection with  Bertalanffy-Richards model}\label{section_bertalanffy}

Analytical solution of Eq. (\ref{EDO_bertalanfy}) is obtained 
when the parameter $D_f$ is keeping fixed during the growth process, for example taking $D_f= D_f^{conv}$. That is reasonable  for $N$ sufficiently large. Thus, one obtains a simpler version of the original approach,  that reveals to be the \textit{Bertallanfy-Richards} growth model \cite{Bertalanffy_model,bertalanffy1960principles,richards1959} whose solution is
\begin{equation}\label{solucao_west}
 m(t) = \left[ \frac{a}{b} + \left(m_0^{1-\beta} - \frac{a}{b} \right) e^{b  (\beta-1) t} \right]^{\frac{1}{1-\beta}},
\end{equation}
 where $a$, $b$ and $\beta$ are  give by~(\ref{Eq_a}), (\ref{Eq_b}), and~(\ref{Eq_beta1}), respectively. 

Fig.~(\ref{Fig_comparacao}) shows a comparison between the original model (evolving $D_f$) and its simplified version (keeping $D_f = D_f^{conv}$ fixed). 
Note that, besides the two dynamics differ during the growth process, 
they converge to the same saturation mass. 

The  Bertalanffy-Richards growth model has been used successfully to describe tumor growth \cite{Guiot2003,model_muddle,Herman2011,Bertalanffy_model,role_growth_rate}.
What is the main point of the connection of our model with the 
Bertalanffy-Richards model is that Eq.~(\ref{EDO_bertalanfy}) and consequently its solution~(\ref{solucao_west}) are obtained by a microscopic approach given by the interaction between cells, and not from a phenomenological (macroscopic) perspective, as has been done in previous works \cite{history1,model_muddle,history2,history3}. 

The solution given by Eq.~(\ref{solucao_west}) has two asymptotic behaviors: saturation or divergence according to the signal of the argument in the exponential of Eq.~(\ref{solucao_west}).
As $\gamma/D_f>1$ (short range interaction regime), then  $(\beta-1)$ is always negative, and consequently the signal of such argument depends only on $b$. 
In fact, this quantity can be written as $b = k_{power} - k$, 
where $k_{power}$ is given by Eq.~(\ref{eq_kpower}). 

If $b<0$ (that is, $k>k_{power}$), then the population growth exponentially and $b$ plays also a role of \textit{growth rate}. 
Otherwise, if $b>0$ (that is, $k<k_{power}$), the population saturates asymptotically. These two accessible phases are limited by a line characterized by $b=0$,  when  the population grows asymptotically as a power law. The Eqs.~(\ref{eq_kpower}) and~(\ref{Eq_power_m}) are obtained in this context.

\begin{figure}[htb!]
\centering
\includegraphics[scale=0.25]{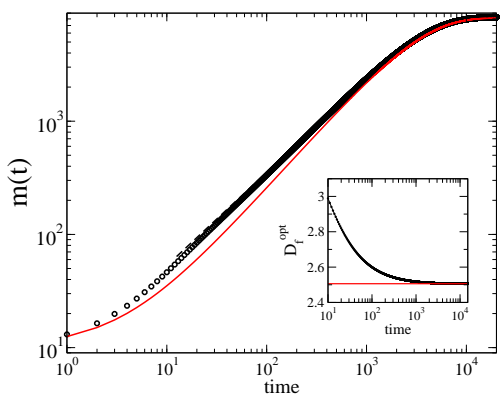}
\caption{ \label{Fig_comparacao} 
Microscopic model prediction of the mass dynamics,  using two approaches: 
1) $D_f$ evolving over time, represented by dots; 
2) keeping the fractal dimension fixed, the red line (special case of the model). 
In both cases were used $J = 0.1$ and $\gamma = 5$. 
The red line show the situation with $D_f$ fixed and numerically identical to $D_f^{conv}$ of the first approach.
The straight dashed line shows a power-law behavior as a guide to the eyes. The red line ($D_f$ constant) is slightly different from the dots curve during the growth process, but they converge to the same saturation value. Inset: population fractal dimension for the two approaches.
In the dots line, the population starts with $D_f^{opt}=D=3$ (compacted form), but  shortly  becomes fractal and converges to a saturation value ($D_f^{opt} \to D_f^{conv}$). }
\end{figure}

\subsection{Connection with Gompertz model}

The Bertalanffy-Richards model is in fact a generalized model, in the sense that it reaches some well know phenomenological models as particular cases.
The relevant quantity in the generalization process is the ratio  $\gamma / D_f$ (or the parameter $\beta$, by Eq.~(\ref{Eq_beta1})). 
For instance,  the \textit{Verhulst model} \cite{Verhulst1845} is reach if $\gamma \ll D_f$ (that is,  $\beta \to 2$). As proposed in \cite{mombach2002,Ribeiro2015b,DOnofrio2009}, this happens when the interaction between cells do not depends on distance. That is, Verhulst model is some kind of mean-field model.

The \textit{Gompertz model} \cite{Gompertz1825} is also a particular case of the Bertalanffy-Richards model, which is reaching when $\gamma \to D_f$ (or $\beta \to 1$). 
It is easy to see how the Gompertz model emerges from the proposed theoretical framework. Defining $\delta=a-b$ and $\alpha=b(1-\beta^{opt}),$ the Bertalanffy-Richards Eq.~(\ref{EDO_bertalanfy}) becomes
\begin{equation}
\frac{dm}{dt}=\delta m^{\beta^{opt}}-\alpha m^{\beta^{opt}}\left(\frac{m^{1-\beta^{opt}}-1}{1-\beta^{opt}}\right).
\label{pregomp}
\end{equation}
If we take the limit $\beta^{opt}\to 1_-:$
\begin{equation}
\frac{dm}{dt}=\delta m-\alpha m \ln{(m)}=-\alpha \ln{\left(\frac{m}{K}\right)}
\label{gomp}
\end{equation}
which is the \emph{Gompertz equation}, with $K\equiv\exp{(\delta/\alpha)}.$ We see that the Gompertz model is recovered at the limit of $ \beta \to 1 $ which corresponds to the optimal fractal dimension $ D_f^{\textrm{opt}} $ be numerically equivalent to the decay coefficient $ \gamma $ (see Eqs.~(\ref{decay},\ref{Eq_beta})).

It is well known the relationship between the tumor malignancy and the fractal dimension of the cancerous cell body. The more malignant the tumor, the greater the fractal dimension \cite{fractals_cancer}. If the condition $D_f = \gamma$ is reasonable, malignancy involves cells with shorter-range interaction when compared to normal cells.

It is also noticed that, for population size big enough, the replication rate decays exponentially, as shown in Fig.~(\ref{fig_mxt}). Our model was formulated based on interactions between the cells which depend just on the distance that separates them. Even considering just this very simple interaction, the model is able to explain very well the success of the phenomenological Gompertz model to describe tumors growth.


\section{Universal Growth Behavior}
\label{sc5}


In  animal growth, conform suggest West {\it et. al.} in Ref. \cite{West2001}, the common pattern observed in the figure~(\ref{fig_universal}) comes from the fact that species allocates energy to growth or to maintenance in the same universal way (see appendix~(\ref{apendice_modelo_west})). 
Moreover, this explanation takes into account the fact that the species obey the \textit{allometric law} (or the \textit{Kleiber's law}), which says that the metabolic rate grows sub-linearly with the mass of the organism \cite{kleiber1932}. However, tumors do not 
necessarily obey such allometric law 
and, furthermore, they present fractal form, instead of space-filling process required by West \textit{et. al.} theory  \cite{West2001, Guiot2006}.
In this way, the explanation based on allocation of energy  is not necessarily suitable for tumors.

Our model, on the other hand,  suggest that such universal growth pattern can come from two principles at microscopic level: competition and self-replication.
In animal growth, the dimensionless mass $\mu$ is related to the ratio between the maintenance energy and the total energy of the organism  (see appendix~(\ref{apendice_modelo_west})). But in our model  this quantity gets another interpretation, although still universal.
In fact, using the relations given by Eqs.~(\ref{Eq_b}), (\ref{Eq_a}), (\ref{eq_kpower}), (\ref{Eq-M}) and (\ref{Eq_mu}),  one can shown that

\begin{equation}
\mu = \frac{ k_{power} - k}{ k_{power} - JI_i}.
\end{equation}
This result means the dimensionless mass is a relationship between the \textit{intrinsic replication rate}  $ k $  and the  \textit{competition rate}  $JI_i$.
While $k$ is constant during the process, the competition rate increase due to the increase of the population size. 
For small times (\textit{i.e.} $\tau \approx 0$), the competition rate is very small compared to the intrinsic replication rate,  resulting in $\mu \approx 1 - k/k_ {power}$. In this case, the tumor keep growing.
However, for $ \tau$ sufficiently large, the competition rate starts to be 
of the same magnitude of the intrinsic replication rate. 
In this case the tumor stops to growth (population saturates), which means $ \mu \to 1$. 
According to the model, as noted in Fig.~(\ref{fig_universal}), no matter the intensity of the competition $J$,  the decay exponent of the competitive interaction $ \gamma $ (and consequently $\beta$), 
or the intrinsic replication rate  $k$, 
all the settings  
collapse into a single universal curve, exactly in the same way that happens to tumor and animal growth.
 Of course, $k$ has to be less than $k_{power}$  to avoid the unrealistic  situation of blow up the population.

This model explain the universal growth pattern  observed in empirical data just using first principles (\textit{i.e.} cell interactions). This is one of the good points of the model. Moreover, this approach also explains that as long as the competition among cells increase, due to the population increase, the replication rate decreases.

As mentioned before, no matter the values of the parameter $\beta$ (given by different  values of $\gamma$ since $ \beta = 2 - \gamma/D_f $), all dynamics collapse in the same universal curve. 
It means that even if an organism does not follow the allometric law ($\beta = 3/4$) it will still follow the universal curve, as is the case of tumors. 

Finally, the microscopic model proposed suggests that the universal similarity between
tumor and animal growth does not necessarily  come  from common biophysical properties. The universal behavior emerges maybe because any biological growth (animal, tumor) are in fact described by the same equation, the Bertalanffy-Richards model~(\ref{EDO_bertalanfy}).  Given a process (biological, physical or whatever) that can be modeled by such a equation, this process will also collapse in the same universal curve.

\section{Conclusions}
\label{sc6}

In this paper, we presented good reasons to consider that our mathematical model is able to explain some empirical evidences about tumor growth.
Our microscopic model describe well enough the form and the growth process of avascular tumors,  
taking into account just few basic principles.
In our prototype, the competition between cells influence their replication rate and the cells can also move in order to minimize this competitive influence from their surrounding cells. 
As were presented, such basic assumptions at microscopic level, conducts to many macroscopic properties that is observed in real tumors. 

The model reproduces, for instance, the exponential growth in early stage  followed by a power law curve for  later times. This patterns is very common in real tumors. The fractal structure, observed in many solid tumors \cite{fractals_cancer,fractalpathology,fractal_medicine}, is also described in our model since the optimal fractal dimension emerges spontaneously, as a consequence of the interrelation between the cells.

Moreover, this model shows that the relation between the intrinsic replication rate and the competition rate of the cells plays the same role of the energy allocation in growing animals. This leads both animal and tumor growth to the same universal behavior. In other words, different biophysical mechanisms can be represented by the same ubiquitous equation, given by Bertalanffy-Richards model. Besides, the universality found for tumor growth is regardless of the values of the parameters of the microscopic model.

In short, we present a robust and powerful model able to describe growth process in general. Regardless of the system concerned, since it was described by the Bertalanffy-Richards model, it will follow the universal growth behavior. In conclusion, we believe that our microscopic model is able to provide a better comprehension of growth patterns and it will can be useful in other fields, involving natural, social and economic contexts.

\section*{Acknowledgements}
We would like to acknowledge the useful and stimulating  discussions with Alexandre Souto Martinez and Fernando Fagundes Ferreira.

\appendix

\section{WBE model for Animal  Growth}\label{apendice_modelo_west}

In this appendix we give a short description of the West model \cite{West2001},  which describes the animal growth considering the Kleiber's  law \cite{kleiber1932}, and the principle of conservation of energy. 
The Kleiber's law says that the metabolic rate $B$ of an organism scales sub-linearly with its body mass, that is $B=B_0 m^{\beta}$, where $B_0$ is a constant and $\beta<1$ is the allometric constant. 
According to the West model,  the total metabolized energy of an organism must be used on maintenance of the already existent cells or in the growth of new cells. That is,
\begin{eqnarray}\label{eq_text}
\text{\small{[Total Metabolic Energy]}}
=\text{\small{[Maintenance]}} + \text{\small{[Growth].}}\nonumber
\end{eqnarray}
It yields the ordinary differential equation:
\begin{equation}
B_0 m^{\beta} =  B_c m + E_c \frac{d m}{dt},
\end{equation}
where $B_c$ is the metabolic rate of a single cell and $E_c$ is the energy necessary to create a new cell. 
The equation above can be write as
\begin{equation}
\frac{dm}{dt} = am^{\beta} - bm,
\label{odeWest}
\end{equation}
where it was defined
\begin{equation}
a \equiv \frac{B_0 m_c}{E_c},
\end{equation}
which is a parameter that does not depend on the species (scale invariant, because it depends only on universal parameters); and 
\begin{equation}
b \equiv \frac{B_c}{E_c}, 
\end{equation}
which depends on the biological species. 

The solution of the Eq. (\ref{odeWest}) is 
\begin{equation}\label{solucao_west0}
m(t) = \left[ \frac{a}{b} + \left(m_0^{1- \beta} - \frac{a}{b} \right) e^{b (\beta - 1) t} \right]^{\frac{1}{  1- \beta}}, 
\end{equation}
where $m_0$ is the initial mass of the organism.

As $\beta<1,$ the solution of Eq.~(\ref{solucao_west0}) converges to 

\begin{equation}\label{Eq-M}
M \equiv m(t\to \infty) = \left( \frac{a}{b} \right)^{\frac{1}{1-\beta}}, 
\end{equation}
which can be interpreted as the maturity mass of the organism. 

One can show that if one plot the quantity

\begin{equation}\label{Eq_mu}
\mu \equiv \left(\frac{m}{M} \right)^{1-\beta} = \frac{\text{Maintenance Energy}}{\text{Total Metabolic Energy}},
\end{equation}
a kind of dimensionless mass, as a function of
\begin{equation}
\tau \equiv -\ln \left( 1 - \left(\frac{m_0}{M}\right)^{1-\beta}  \right) + \frac{a(1-\beta)}{M^{1-\beta}} t,  
\end{equation}
a kind of dimensionless time,
many kinds of animals (birds, mammals, fish) and also tumors collapse in the same universal curve ($\mu = 1- e^{-\tau}$), as it was showed in the figure~(\ref{fig_universal}). 







%

\end{document}